\documentclass[pra,twocolumn,showpacs,superscriptaddress]{revtex4}%
\usepackage[dvips]{graphics,color}
\usepackage{amsmath}
\usepackage{amssymb}
\usepackage{graphicx}
\usepackage{amsfonts}
\usepackage{epsfig}
\usepackage{CJK}%
\providecommand{\U}[1]{\protect\rule{.1in}{.1in}}

\begin{document}
\title{Extended JC-Dicke model for a two-component atomic BEC inside a cavity}
\author{Yong Li}
\affiliation{Department of Physics and Center of Theoretical and Computational Physics, The
University of Hong Kong, Pokfulam Road, Hong Kong, China}
\author{Peng Zhang}
\affiliation{ERATO, JST, Macroscopic Quantum Control Project, Hongo, Bunkyo-Ku, Tokyo
113-8656, Japan}
\author{Z. D. Wang}
\affiliation{Department of Physics and Center of Theoretical and Computational Physics, The
University of Hong Kong, Pokfulam Road, Hong Kong, China}

\begin{abstract}
We consider a trapped two-component atomic Bose-Einstein condensate (BEC),
where each atom with three energy-levels is coupled to an optical cavity field
and an external classical optical field as well as a microwave field to form
the so-called $\Delta$-type configuration. After adiabatically eliminating the
atomic excited state, an extended JC-Dicke model is derived under the
rotating-wave approximation. The scaled ground-state energy and the phase
diagram of this model Hamiltonian are investigated in the framework of
mean-field approach. A new phase transition is revealed when the amplitude of
microwave field changes its sign.

\end{abstract}

\maketitle


\section{Introduction}

The so-called Dicke model \cite{Dicke:1954} describes the
interaction of a large number of two-level systems (e.g., atoms)
with a single optical mode. Since the effective light-matter
coupling strength is dependent on the number of atoms $N$
($\propto\sqrt{N}$), a sufficiently large $N$
would lead to a classical phase
transition~\cite{Hepp:1973,Wang:1973} (at finite temperatures) from
the normal state, which corresponds to the atomic ground state
associated with the vacuum state of the optical mode, to the
superradiant-phase state, where the phenomenon of superradiance
occurs with the finite scaled mean numbers of both photons and
excited-state atoms. Recently, exploration of quantum phase
transitions in the Dicke model at zero temperature
has attracted significant attentions
\cite{Emary:2003,Lambert,Hou:2004,Emary:2004,Lee:2004,Li:2006,Vidal:2006,
Alcalde:2007,Dimer:2007,Tolkunov:2007,Chen:2007a,Chen:2007b,Chen:2008,
Goto:2008,Tsyplyatyev:2008,QHChen:2008,Alcalde:2009,Larson:2009,Huang:2009}.
The drastic change across the critical point due to a qualitative change of
the ground state of the Dicke model has been investigated, in the frameworks
of the scaled ground state energy, macroscopic atomic excited-state
population, quantum entanglement, Berry phase, quantum chaos and so on.

Apart from the standard Dicke model, several more generalized Dicke
models have also been proposed and studied
\cite{Emary:2004,Lee:2004,Li:2006,Tolkunov:2007,Chen:2007b,Chen:2008,
Goto:2008,Tsyplyatyev:2008,Alcalde:2009,Larson:2009}. In some of
them \cite{Chen:2007b,Chen:2008}, the free atoms in the standard
Dicke models are replaced by the atomic Bose-Einstein condensate
(BEC). A BEC describes a collective quantum state of a large number
of atoms and may be used to generate a macroscopic quantum object
with a longer lifetime compared to the free atoms. It has attracted
much interest to combine the BEC with optical cavity in the strong
coupling regime
\cite{Ottl:2006,Brennecke:2007,Brennecke:2008,Murch:2008}. On the
other hand, comparing with the case of free atoms in the standard
Dicke model, there is an additional atom-atom collision interaction
in the BEC, which could lead to the change of critical phenomenon
\cite{Chen:2007b}. More recently, under some approximations,
Chen~\textit{et al.}~\cite{Chen:2008} proposed an extended Dicke
model based on a two-component atomic BEC in an optical cavity with
the atoms coupled to both the quantized optical cavity filed and an
external classical driving optical field, where an interesting phase
diagram covering phenomena from quantum optics to atomic BEC is
addressed.

In this paper, we consider a trapped two-component atomic BEC where the two
condensated states and an ancillary excited state of each atom form a
three-level $\Delta$-type configuration via the couplings to the optical
cavity and an externally controlled classical optical field as well as an
external microwave field. Such a three-level configuration can be reduced to a
two-level configuration (i.e. the two condensated states) after adiabatically
eliminating the atomic excited state under the large sigle-photon detuning
condition. Under the so-called rotating-wave approximation (RWA), we derive an
effective, extended JC-type~\cite{JC} Dicke model for such a two-component
BEC. The phase diagram for the derived extended JC-Dicke model is also
investigated in detail.

\section{Model and Hamiltonian}

The setup of the JC-Dicke model under consideration is depicted in
Fig.~\ref{setup02}. An optically-trapped Rb atomic BEC under the two-mode
approximation with the atomic states $5^{2}S_{1/2}$ $\left\vert F=1,m_{f}%
=-1\right\rangle $ (ground state $\left\vert 1\right\rangle $) and $\left\vert
F=1,m_{f}=0\right\rangle $ (metastable state $\left\vert 2\right\rangle $) is
placed in a single-mode quantized optical cavity. These two states of the
atomic BEC are coupled via both an external microwave field and a two-photon
process \cite{two-photon BEC} mediated by an ancillary excited state
$\left\vert 3\right\rangle $ (from $5^{2}P_{3/2}$), where the single-photon
transition between $\left\vert 3\right\rangle $ and the ground state
$\left\vert 1\right\rangle $ (or the metastable state $\left\vert
2\right\rangle $) is coupled to the quantized optical cavity field (or the
external classical optical field). Here we assume that both the corresponding
single-photon detunings are large and the corresponding two-photon detuning is
very small. For such a case of two-photon Raman process \cite{Raman}, the
ancillary excited state can be adiabatically eliminated and the effective
Hamiltonian of the two-component BEC reads ($\hbar=1$ hereafter)
\begin{equation}
H_{\mathrm{eff}}=H_{at}+H_{ph}+H_{at-ph}+H_{at-cl}+H_{at-at}\label{Ham-1}%
\end{equation}
with
\begin{subequations}
\begin{align}
H_{at} &  =\nu_{1}c_{1}^{\dagger}c_{1}+(\nu_{2}+\omega_{2}-\omega_{1}%
)c_{2}^{\dagger}c_{2},\\
H_{ph} &  =\omega a^{\dagger}a,\\
H_{at-op} &  =\lambda_{\mathrm{eff}}e^{i\omega_{\mathrm{cl}}t}c_{2}^{\dagger
}c_{1}a+h.c.,\\
H_{at-mw} &  =\Omega e^{-i\omega_{\mathrm{mw}}t}c_{2}^{\dagger}c_{1}+h.c.,\\
H_{at-at} &  =\frac{\eta_{1}}{2}c_{1}^{\dagger}c_{1}^{\dagger}c_{1}c_{1}%
+\frac{\eta_{2}}{2}c_{2}^{\dagger}c_{2}^{\dagger}c_{2}c_{2}+\eta_{12}%
c_{1}^{\dagger}c_{1}c_{2}^{\dagger}c_{2}%
\end{align}
denoting the energies of the free atoms in the BEC, the free cavity field, the
reduced effective interaction between the BEC with the optical fields, the
interaction of the BEC with the microwave field, and the atom-atom collision
interaction of the BEC, respectively. Here, $c_{1,2}$ ($c_{1,2}^{\dagger}$)
are the annihilation (creation) bosonic operators for $\left\vert
1\right\rangle $ and $\left\vert 2\right\rangle $, respectively; $\omega_{i}$
($i=1,2,3$) is the corresponding internal level-energy for atomic state
$\left\vert i\right\rangle $; $\nu_{l}=\int d^{3}\mathbf{r}\phi_{l}^{\ast
}(\mathbf{r})[-\nabla^{2}/2m+V(\mathbf{r})]\phi_{l}(\mathbf{r})$ ($l=1,2$) is
the trapped frequency for the states $\left\vert 1\right\rangle $ and
$\left\vert 2\right\rangle $ with $V(\mathbf{r})$ being the trapped potential,
$m$ the atomic mass, and $\phi_{l}(\mathbf{r})$ the corresponding condensate
wavefunction; $a$ ($a^{\dagger}$) is the annihilation (creation) operator of
the cavity mode with the frequency $\omega$; $\lambda_{\mathrm{eff}}%
=g_{13}\Omega_{23}/\Delta$ is the reduced effective coupling strength for
two-photon Raman process, where $g_{13}$ ($\Omega_{23}$) is the corresponding
coupling strength between the BEC and the quantized cavity field (classical
optical field), $\Delta$ is the large single-photon detuning: $\Delta
\equiv\omega_{3}-$ $(\nu_{2}+\omega_{2})-\omega_{\mathrm{cl}}\gg
\{g_{13},\Omega_{23}\}$ with $\omega_{\mathrm{{cl}}}$ the frequency of the
classical optical field; $\Omega$ is the corresponding coupling strength
between the BEC and the microwave field (with the frequency $\omega
_{\mathrm{{mw}}}$); $\eta_{l}=(4\pi\rho_{l}/m)\int d^{3}\mathbf{r}\left\vert
\phi_{l}(\mathbf{r})\right\vert ^{4}$ and $\eta_{12}=(4\pi\rho_{12}/m)\int
d^{3}\mathbf{r}\left\vert \phi_{1}^{\ast}(\mathbf{r})\phi_{2}(\mathbf{r}%
)\right\vert ^{2}$ with $\rho_{l}$ and $\rho_{12}$ ($=\rho_{21}$) the
intraspecies and the interspecies $s$-wave scattering lengths, respectively.
It is remarked that the RWA has been used for all optical/microwave fields
coupling to the atomic BEC.

\begin{figure}[th]
\includegraphics[width=8cm]{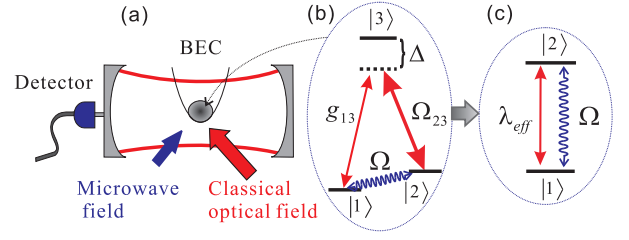}
\caption{(Color online) (a) Schematic diagram of the experimental setup for a
trapped BEC of $^{87}$Rb atoms. (b) The internal atomic level configuration of
BEC: The $\left\vert 1\right\rangle $$\leftrightarrow$$\left\vert
2\right\rangle $ transition of the BEC atoms is coupled to the external
microwave field with the coupling strength $\Omega$; The $\left\vert
1\right\rangle $$\leftrightarrow$$\left\vert 3\right\rangle $ and $\left\vert
2\right\rangle $$\leftrightarrow$$\left\vert 3\right\rangle $ transitions are
coupled largely-detuned to the optical cavity and the external classical
optical field with the corresponding coupling strengths $g_{13}$ and
$\Omega_{23}$, respectively. (c) When the single-photon detuning $\Delta$ is
large (the corresponding two-photon detuning is assumed to be small), a
reduced two-level configuration can be obtained after eliminating the
ancillary excited state $\left\vert 3\right\rangle $. $\lambda_{\mathrm{eff}}$
is the effective coupling strength for the two-photon Raman process.}%
\label{setup02}
\end{figure}

By using the Schwinger relations
\end{subequations}
\begin{subequations}
\begin{align}
J_{+} &  =c_{2}^{\dagger}c_{1},\text{ \ \ }J_{-}=c_{1}^{\dagger}c_{2},\\
J_{z} &  =\frac{c_{2}^{\dagger}c_{2}-c_{1}^{\dagger}c_{1}}{2},
\end{align}
which fulfill
\end{subequations}
\begin{equation}
\left[  J_{+},J_{-}\right]  =2J_{z},\text{ \ }\left[  J_{z},J_{\pm}\right]
=\pm J_{\pm},
\end{equation}
the Hamiltonian (\ref{Ham-1}) can be written as
\begin{align}
H_{\mathrm{eff}} &  =\omega a^{\dagger}a+\omega_{0}J_{z}+\frac{\eta}{N}%
J_{z}^{2}\nonumber\\
&  +[(\Omega e^{-i\omega_{\mathrm{mw}}t}+\frac{\lambda}{\sqrt{N}}%
e^{i\omega_{\mathrm{cl}}t}a)J_{+}+h.c.]+\text{const},\label{Ham-2}%
\end{align}
where
\begin{equation}
N=c_{2}^{\dagger}c_{2}+c_{1}^{\dagger}c_{1}%
\end{equation}
is the number of the atoms,
\begin{subequations}
\begin{align}
\omega_{0} &  =\nu_{2}+\omega_{2}-\nu_{1}-\omega_{1}+\frac{N-1}{2}\left(
\eta_{2}-\eta_{1}\right)  ,\\
\eta &  =(\frac{\eta_{1}+\eta_{2}}{2}-\eta_{12})N,\text{ }\\
\lambda &  =\lambda_{\mathrm{eff}}\sqrt{N},
\end{align}
and the constant term
\end{subequations}
\[
\text{const}=\frac{N}{2}[(\nu_{2}+\omega_{2}-\omega_{1}-\frac{\eta_{2}}%
{2}+\gamma_{2}N)+(\nu_{1}-\frac{\eta_{1}}{2}+\gamma_{1}N)]
\]
can be neglected in the following consideration.

For $\Omega\neq0$, to eliminate the time-dependence in Hamiltonian
(\ref{Ham-2}), we perform a unitary transformation $U=\exp[-i\omega
_{\mathrm{{mw}}}J_{z}t-i(\omega_{\mathrm{{mw}}}+\omega_{\mathrm{{cl}}%
})a^{\dagger}at]$
and obtain an effective Hamiltonian
\begin{align}
H &  =\omega_{a}a^{\dagger}a+\omega_{b}J_{z}+\frac{\eta}{N}J_{z}%
^{2}\nonumber\\
&  +(\frac{\lambda}{\sqrt{N}}aJ_{+}+\Omega J_{+}+h.c.),\label{Ham-3}%
\end{align}
where $\omega_{a}=\left(  \omega-\omega_{\mathrm{mw}}-\omega_{\mathrm{cl}%
}\right)  $ and $\omega_{b}=\left(  \omega_{0}-\omega_{\mathrm{mw}}\right)  $
are the effective frequencies in the rotating frame, in which $H$ is
independent of time. To the best of our knowledge, Hamiltonian (\ref{Ham-3})
appears to be a new one in literatures, and thus we call it as the extended
JC-Dicke model.

\section{Mean-field ground state energy}

We now look into the ground-state properties of Hamiltonian (\ref{Ham-3}) and
the corresponding quantum phases as well as their transitions. We here
consider the case of positive $\omega_{a}$, where the stable ground state is
anticipated for Hamiltonian (\ref{Ham-3}). By using the Holstein-Primakoff
transformation \cite{HP}
\begin{align}
J_{+}  &  =b^{\dagger}\sqrt{N-b^{\dagger}b},\text{\ }J_{-}=\sqrt{N-b^{\dagger
}b}b,\nonumber\\
J_{z}  &  =b^{\dagger}b-\frac{N}{2}%
\end{align}
with $[b,b^{\dagger}]=1$, the effective Hamiltonian reads
\begin{align}
H  &  =\omega_{a}a^{\dagger}a+\omega_{b}(b^{\dagger}b-\frac{N}{2})+\frac{\eta
}{N}(b^{\dagger}b-\frac{N}{2})^{2}\nonumber\\
&  +[(\lambda a+\Omega\sqrt{N})b^{\dagger}\sqrt{1-\frac{b^{\dagger}b}{N}%
}+h.c.]. \label{Ham-4}%
\end{align}

Similar to that used in the standard Dicke model \cite{Emary:2003}, we here
introduce the displacements for the two shifting boson operators as
$c^{\dagger}=a^{\dagger}-\sqrt{N}\alpha^{\ast}$ and $d^{\dagger}=b^{\dagger
}+\sqrt{N}\beta^{\ast}$ with the complex displacement parameters $\alpha$ and
$\beta$ describing the scaled collective behaviors of both the atoms and the
photons \cite{Emary:2003,Li:2006,Chen:2007b,Chen:2008}. In fact, the current
method of introducing the displacements is equivalent to the mean field
approach. In this framework, it is clear that $0\leq\left\vert \beta
\right\vert \leq1$.

After expanding the terms in the square in (\ref{Ham-4}), the scaled
Hamiltonian can be written up to the order of $N^{-1}$ as
\begin{equation}
H/N=H_{0}+N^{-1/2}H_{1}+N^{-1}H_{2},
\end{equation}
where
\begin{align}
H_{0} &  =\omega_{a}\alpha^{\ast}\alpha+\omega_{b}(\beta^{\ast}\beta-\frac
{1}{2})+\eta(\beta^{\ast}\beta-\frac{1}{2})^{2}\nonumber\\
&  -[\left(  \lambda\alpha+\Omega\right)  \beta^{\ast}\sqrt{1-\beta^{\ast
}\beta}+c.c.]
\end{align}
denotes the scaled constant energy, and $H_{1,2}$ denote the linear
and bilinear terms, respectively. It is noted that $H_{0,1,2}$ are
independent of the number of atoms $N$.

The scaled ground-state energy is just given by the scaled constant energy in
the Hamiltonian
\begin{equation}
E_{g}^{N}(\alpha,\beta)\equiv\frac{E_{g}(\alpha,\beta)}{N}= H_{0},
\label{scaled energy}%
\end{equation}
where the displacements $\alpha$ and $\beta$ should be determined from the
equilibrium condition
\begin{subequations}
\begin{align}
\partial\lbrack E_{g}(\alpha,\beta)/N]/\partial\alpha^{\ast}  &
=0,\label{partial-alpha}\\
\partial\lbrack E_{g}(\alpha,\beta)/N]/\partial\beta^{\ast}  &  =0.
\end{align}
After some derivations, we find that $\alpha$ is given by
\end{subequations}
\begin{equation}
\alpha=\frac{\lambda^{\ast}}{\omega_{a}}\beta\sqrt{1-\beta^{\ast}\beta}.
\label{alpha-solve}%
\end{equation}
and $\beta$ satisfies
\begin{align}
\Omega\sqrt{1-\beta^{\ast}\beta}=\beta\lbrack &  \omega_{b}+w\left(
2\beta^{\ast}\beta-1\right) \nonumber\\
&  +(\frac{\Omega\beta^{\ast}}{2\sqrt{1-\beta^{\ast}\beta}}+c.c.)],
\label{belta-solve}%
\end{align}
where $w=\eta+\left\vert \lambda\right\vert ^{2}/\omega_{a}$.

From the above equation (\ref{belta-solve}), it is obvious that $\Omega/\beta$
should be real since all of the parameters, except for the coupling strengths
$\lambda$ and $\Omega$, are real. Here, we assume the case of real $\Omega$
for simplicity \cite{Note:Rabi frequency}. That means, $\beta$ should also be
real: $-1\leq\beta\leq1$ and satisfy
\begin{equation}
0=\omega_{b}\beta\sqrt{1-\beta^{2}}+\Omega\left(  2\beta^{2}-1\right)
+w\beta\left(  2\beta^{2}-1\right)  \sqrt{1-\beta^{2}}. \label{belta-solve2}%
\end{equation}


For a general real $\Omega$, the scaled ground state energy in
Eq.~(\ref{scaled energy}) is given by the displacement $\beta$ (by using
Eq.~(\ref{alpha-solve}) to eliminate the displacement $\alpha$) as
\begin{equation}
E_{g}^{N}(\beta)=\omega_{b}(\beta^{2}-\frac{1}{2})-2\Omega\beta\sqrt
{1-\beta^{2}}+w(\beta^{2}-\frac{1}{2})^{2}. \label{scaled energy-2}%
\end{equation}
The displacement $\beta$ is the non-trivial real solution for
equation~(\ref{belta-solve2}). In general, there are more than one real
solutions for Eq.~(\ref{belta-solve2}), only the one that leads to the minimal
scaled ground state energy should be chosen.

We would like to remark that the displacements determined from the
equilibrium equations could just make the linear term $H_{1}$ be
$0$. At the same time, the bilinear term $H_{2}$ makes no
contribution to the scaled ground state energy. Therefore, the exact
forms of $H_{1,2}$ are not needed in our analysis, and thus only the
constant term $H_{0}$ determines fully the scaled ground state
energy of the current system.
%
\begin{figure}[th]
\includegraphics[width=7cm]{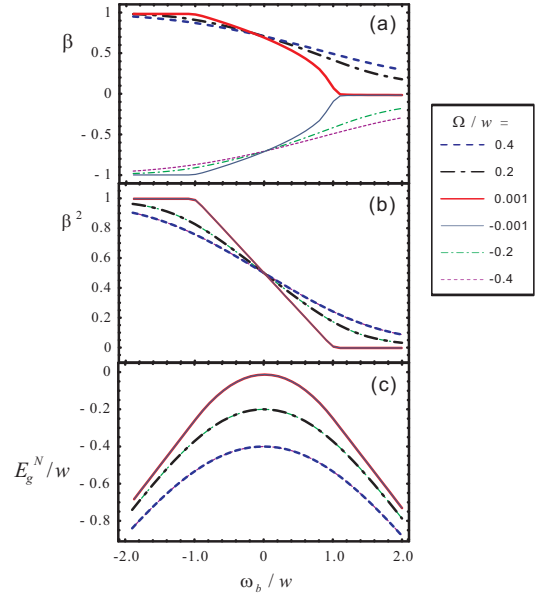}
\caption{(Color online) (a) The atomic displacement $\beta$ for the
ground state, (b) the square of atomic displacement $\beta^{2}$ for the
ground state (or the scaled magnetization plus $1/2$: $m+1/2$), and (c)
the scaled ground state energy $E_{g}^{N}$ versus $\omega_{b}$ for
different coupling strength $\Omega$. The
energy and frequencies are in units of $w$ ($w>0$).}%
\label{fig1}%
\end{figure}

\section{Numerical results and analysis}

In this section, we focus on the numerical calculations of the scaled ground
state energy, e.g., $E_{g}^{N}(\beta)$ in Eq.~(\ref{scaled energy-2}), and the
corresponding displacement $\beta$ that makes $E_{g}^{N}(\beta)$ minimal. The
minimum $E_{g}^{N}(\beta)$ (as well as the corresponding $\beta$) is
determined by the three parameters: $\omega_{b}$, $\Omega$, and $w$.

Figure~\ref{fig1} plots $\beta$, $\beta^{2}$, and $E_{g}^{N}(\beta
)$\ (corresponding to the ground state of the JC Dicke model) as a
function of $\omega_{b}$ for several values of $\Omega$. All the
energies/frequencies in Fig.~\ref{fig1} are in units of the positive
$w$. We mentioned \textquotedblleft positive $w$\textquotedblright\
here since $w$ may be positive or negative, while the negative $w$
will lead to different results. Seen from Fig.~\ref{fig1}, a
second-order normal-superradiant phase transition at
$\omega_{b}/w=1$ should be expected in the absence of microwave
field ($\Omega=0$), as in the standard Dicke
model~\cite{Emary:2003};
while in the presence of microwave field ($\Omega\neq0$), the normal phase and
the corresponding transition disappear.
%
\begin{figure}[th]
\includegraphics[width=6cm]{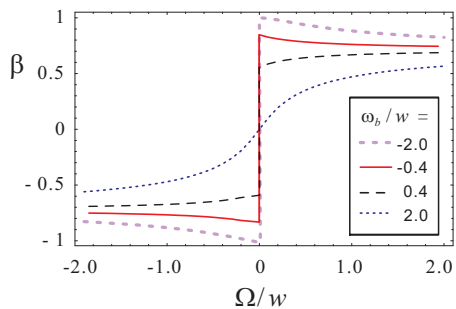}
\caption{(Color online) The atomic displacement $\beta$ for the ground
state versus the coupling strength $\Omega$ for different $\omega_{b}$.
The frequencies are
in units of $w$ ($w>0$).}%
\label{fig2}%
\end{figure}
\begin{figure}[th]
\includegraphics[width=9cm]{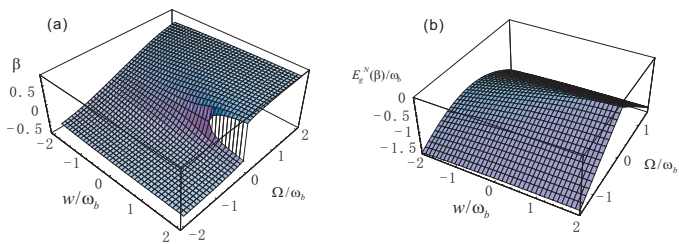}
\caption{(Color online) (a) The atomic displacement $\beta$ for the
ground state and (b) the scaled ground state energy $E_{g}^{N}$ versus
$\Omega$ and $w$. The
frequencies/energies are in units of $\omega_{b}$ (here we keep $\omega_{b}%
>0$).}%
\label{fig3}%
\end{figure}

The displacement $\beta$ for the ground state is plotted as a
function of $\Omega$ (in units of positive $w$) for several values
of $\omega_{b}$ in Fig.~\ref{fig2}. For a positive/negative
$\Omega$, the corresponding $\beta$ is also positive/negative. At
the point of $\Omega\rightarrow0$, the ground-state $\beta$ may have
a jump.
In order to look the possible transition phenomenon at the point of
$\Omega\rightarrow0$, we plot the ground-state $\beta$ and $E_{g}^{N}(\beta)$
as the function of the parameters $\Omega$ and $w$ (in units of positive
$\omega_{b}$) in the 3-dimensional (3D) Fig.~\ref{fig3}. In experiments, the
parameters $\Omega$ and $w$ are controllable, e.g., the former one can be
easily controlled by changing the strength of the microwave field, the latter
one can be controlled by adjusting the atom-atom interaction interactions by
the magnetic-field or/and optical Feshbach resonance techniques
\cite{m-Feshbach,o-Feshbach,Zhang:2009}. From Fig.~\ref{fig3}(a), it is clear
that the displacement $\beta$ has a jump when $w>\omega_{b}$ at the point of
$\Omega\rightarrow0$. Notably, the scaled ground-state energy is always
continuous at the jump point, but its first derivative with respect to
$\Omega$ does not, which implies a new kind of the first order phase
transition at the parameter point whenever $\Omega$ changes its sign. Making
the replacement: $\Omega\rightarrow-\Omega$, we can find the corresponding
replacements $\beta\rightarrow-\beta$ and $E_{g}^{N}(\beta)\rightarrow
E_{g}^{N}(-\beta)=E_{g}^{N}(\beta)$ according to Eqs.~(\ref{belta-solve2}) and
(\ref{scaled energy-2}), just as seen in Fig.~\ref{fig3}.

\begin{figure}[th]
\includegraphics[width=8cm]{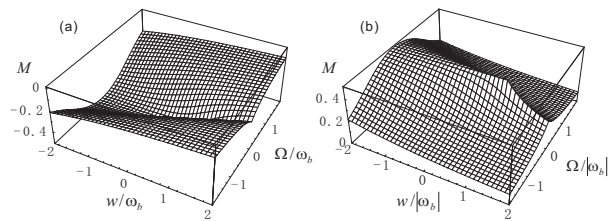}
\caption{(Color online) The atomic scaled magnetization $M$ versus
$\Omega$ and $w$. The parameters are in units of $|\omega_{b}|$ for
positive (a) and negative (b) $\omega_{b}$.}
\label{fig4}%
\end{figure}
\begin{figure}[th]
\includegraphics[width=8cm]{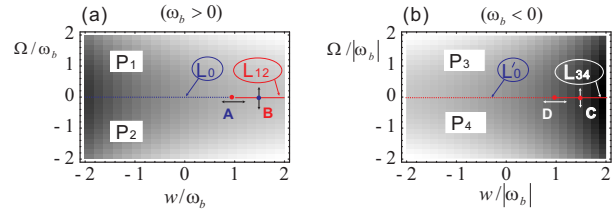}\caption{(Color online) The $\Omega$-$w$
phase diagram for (a) the positive $\omega_{b}$ and (b) negative $\omega_{b}
$.}%
\label{fig5}%
\end{figure}

We may define the scaled 'magnetization' as $M\equiv\langle J_{z}
\rangle/N =\beta^{2}-1/2$ for the ground state. A positive
magnetization $M$ means the atomic inversion population: more atoms
stay in the upper state $\left\vert 2\right\rangle $ than those in
$\left\vert 1\right\rangle $, while a negative $M$ means the opposite
case. Especially, $M=-1/2$ corresponds to the normal phase: all atoms
stay in the lower states. Figure~\ref{fig4} plots the magnetization
against the parameters $\Omega$ and $w$, where Figs.~\ref{fig4}(a) and
(b) correspond respectively to the positive and negative $\omega_{b}$.
From Fig.~\ref{fig4}, we may define phases $P_{1,2,3,4}$, as denoted in
Fig.~\ref{fig5} and Table~\ref{table-1}. The line $L_{0}$ of normal
phase and $L_{12}$ of the superradiant phase, appearing only in the
absence of microwave field, are the board lines of the phases $P_{1,2}$
in Fig.~\ref{fig5}(a); so are $L_{0}^{\prime}$ and $L_{34}$ the board
lines of $P_{3,4}$ in Fig.~\ref{fig5}(b). Table~\ref{table-1} shows the
differences of relevant quantities for all phases in
Fig.~\ref{fig5}~\cite{note}.

We now address the points/line of quantum phase transition in the
phase diagram. Point $A$ in Fig.~\ref{fig5}(a) denotes the
well-known normal-superradiant phase transition (along the
transverse arrow) in the standard JC Dicke model, while point $D$
represents a similar one (corresponding to point $A$) for the case
with the atomic near-inversion population when $\omega_{b}$ is
negative. The line $L_{12}$ corresponds to the phase transition line
segment from $P_{1}$ to $P_{2}$, which is of the first-order, as
indicated above. Intuitively, the phases $P_{1}$ and $P_{2}$ may be
viewed as the para- and dia- ``magnetic" phases, because $\frac
{\partial M}{\partial\Omega}>0$ and $<0$ for $P_{1}$ and $P_{2}$,
respectively. Correspondingly, the line segment of $L_{34}$ is the
first-order phase transition line that separates $P_{3}$ and
$P_{4}$, which may be viewed respectively as the dia- and para-
``magnetic" phases.
%
\begin{table}[h]
\caption{Relevant quantities for different phases $P_{1,2,3,4}$ in the ground
state.}%
\label{table-1}
\begin{tabular}
[c]{lllll}\hline\hline
quantity \ \  & P1 & P2 & P3 & P4\\\hline\hline
$\omega_{b}$ \ \  & $>0$ \ \ \ \ \ \  & $>0$ \ \ \ \ \ \  & $<0$
\ \ \ \ \ \  & $<0$\\
$\Omega$ \ \  & $>0$ \ \ \ \ \ \  & $<0$ \ \ \ \ \ \  & $>0$ \ \ \ \ \ \  &
$<0$\\\hline
$\beta$ \ \  & $(0,\frac{\sqrt{2}}{2})$ \ \  & $(-\frac{\sqrt{2}}{2},0)$
\ \  & $(\frac{\sqrt{2}}{2},1)$ \ \  & $(-1,-\frac{\sqrt{2}}{2})$\\
$M$ \ \  & $(-\frac{1}{2},0)$ \ \ \ \ \ \  & $(-\frac{1}{2},0)$ \ \ \ \ \ \  &
$(0,\frac{1}{2})$ \ \ \ \ \ \  & $(0,\frac{1}{2})$\\\hline
$\frac{\partial E_{g}^{N}}{\partial\Omega}$ \ \ \ \  & $<0$ & $>0$ \ \  & $<0$
\ \  & $>0$\\
$\frac{\partial M}{\partial w}$ \ \  & $>0$ & $>0$ & $<0$ & $<0$\\
$\frac{\partial M}{\partial\Omega}$ \ \  & $>0$ & $<0$ & $<0$ & $>0$%
\\\hline\hline
\end{tabular}
\end{table}

It is worth pointing that we have neglected the anti-resonant terms
and the corresponding $\hat{A}^{2}$ terms ($\hat{A}$ is the vector
potential of the optical field) in the original Hamiltonian. In the
standard Dicke model, it was pointed out that the anti-resonant
terms would also bring an un-neglectful influence near the critical
point and lead to the modification of the result of the quantum
phase transition~\cite{Duncan:1974,Emary:2003}. It was also pointed
out that the quantum phase transition in the standard Dicke model
happens in the effective ultra-strong matter-light coupling regime,
where the $\hat{A}^{2}$ term could also become very strong and may
not be omitted. Notably, if the effect of the $\hat{A}^{2}$ term
were not neglected, the quantum phase transition would be impossible
to happen \cite{Rzazewski:1975} in the standard Dicke model. In
order to obtain the quantum phase transition in the Dicke model, it
was proposed in Ref.~\cite{Dimer:2007} to get an effective Dicke
model which does not include the $\hat{A}^{2}$ term. In the current
scheme, similar to Ref.~\cite{Dimer:2007}, we have obtained an
effective extended JC-Dicke model that does not have the
anti-resonant terms and the $\hat{A}^{2}$ terms. In our original
Hamiltonian, it is assumed that the matter-light couplings are much
smaller than the corresponding atomic transition/optical carrier
frequencies, then it is safe to neglect the anti-resonant terms and
the $\hat{A}^{2}$ terms. After the unitary transformation, the
time-independent effective Hamiltonian may lead to the quantum phase
transition when the matter-light coupling is comparable to the
effective carrier frequencies.

\section{Conclusion}

In conclusion, we have derived an extended JC-Dicke model for a two-component
BEC coupled to the quantized optical cavity and the external classical optical
field as well as a microwave field. The scaled ground-state energy and the
phase diagram of this model Hamiltonian have been investigated in the
framework of mean-field approach. A new first-order phase transition has also
been revealed when the amplitude of micromave field changes its sign.

\begin{acknowledgments}
We thank Ming-Yong Ye, Zi-Jian Yao, Gang Chen, and Zheng-Yuan Xue for helpful
discussions. This work was supported by the RGC of Hong Kong under Grant No.
HKU7051/06P, the URC fund of HKU, and the State Key Program for Basic Research
of China (No. 2006CB921800).
\end{acknowledgments}


\begin{thebibliography}{99}
\bibitem {Note:Rabi frequency}Usually, when an atomic transition is coupled to
only one optical/microwave field, the phase factor of the coupling strength
(i.e., $\Omega$ here) can be absorbed into the atomic state by redefining the
state and thus the coupling strength can be taken to be always
\textquotedblleft positive\textquotedblright. However, it is not the case if a
given atomic transition is coupled to two independent (optical/microwave)
fields (or identically coupled to two fields as considered here). In this
case, one should not take the both coupling strengths to be always
\textquotedblleft positive\textquotedblright. In this paper, we consider the
coupling strength $\Omega$ to be a real value (either positive or negative)
for the sake of simplicity (here an opposite sign of $\Omega$ may be viewed to
denote the opposite direction of the \textquotedblleft external field-$\Omega
$\textquotedblright, as seen from Eq.~(\ref{Ham-3})).
\bibitem {note}We here do not adopt the scenario used in Ref.~\cite{Chen:2008}
to address the Mott-superfluid phase transition. It is because such a phase
transition may be well defined in the mean-field framework only for
$\omega_{b}=0$.
\end{thebibliography}
\end{document}